# Temperature variability over the Po Valley, Italy, according to radiosounding data

Boyan Hristozov Petkov

Institute of Atmospheric Sciences and Climate (ISAC), Italian National Research Council (CNR), Via Gobetti 101, I-40129 Bologna, Italy (e-mail: b.petkov@isac.cnr.it)

**Abstract**

Temperature variations registered above the southeast part of the Po Valley, Italy have been examined by applying the principal component analysis of radiosoung profiles recorded during the period from 1987 to 2010. Two data sets, considered to describe intra- and inter-annual oscillations, respectively were extracted from the measurements data and the results show that both types of fluctuations can be projected onto four empirical orthogonal functions (EOF), interpreted as vertical distributions of oscillation amplitudes and four uncorrelated time series that represent the evolution of corresponding EOFs. It was found that intra-annual oscillations composed of periods between 30 and 120 days, together with inter-annual variations of 1- to 7-year period contribute to the highest extent (about 70%) the temperature oscillations up to 20 km, changing in both cases the phase in the tropopausal region. The other three EOFs indicate prevailing weight of the oscillations in the upper troposphere – low stratosphere region and are characterized by longer periods in both types of fluctuations. The intra-annual variations can be accounted for an interaction between Madden-Julian and Arctic oscillations, while the spectral features of inter-annual fluctuations could be associated with those of Quasi Biennial, El Niño and North Atlantic global oscillations.

## 1. Introduction

The air temperature at different altitude levels is strongly impacted by dynamical processes and, as a result, it is a subject of variations pertaining to large frequency diapason [1, 2, 3]. Such variations could be associated with the numerous patterns of oscillations observed in various atmospheric and oceanic parameters, like Madden-Julian Oscillations (MJO) [4, 5], Arctic Oscillations (AO) [6, 7], Quasi-biennial Oscillations (QBO) [8], El Niño Southern Oscillations (ENSO) [9, 10, 11], North Atlantic Oscillations (NAO) [10, 12]. Despite that the major part of these variations are generated in the equatorial and tropical zones they frequently expand to the extra-tropical regions affecting mid-latitude atmosphere. For instance, it was found that ENSO episodes, which are associated with see surface temperature of the tropical Pacific strongly impact extra-tropical atmosphere [13] and such a propagation is more effective at middle latitudes in the Northern Hemisphere, where ENSO wave-like anomalies are observed up to 35 – 40 km altitude [11]. These atmospheric oscillations are consisted of periods occupying comparatively large time scale starting from intraseasonal MJO composed of periods between 40 and 50 days [4] and AO that has a very broad time spectrum ranging from weekly to seasonal and even to decadal scales [14]. Further, QBO is a mode with variable period averaging approximately 28 months [8] and, ENSO and NAO oscillations consisted of periods ranging from 2 to 10 years with peaks at 3 and 7 years for ENSO [10] and 2.5 and 6–10 years for NAO [12]. In addition, the interaction among these oscillations or between one of them with the annual cycle could produce amplitude modulations or variations with intermediate frequencies [8, 14, 15, 16].

Since the temperature is an important parameter determining to high extent the climate, the studding of its variations in the atmosphere is closely related to the testing and improvement of climatic models [17, 18, 19, 20]. The present study is aimed to examine the altitude-temporal features of temperature variations observed over the southeast part of the Po Valley, Italy, by analysing the radiosoung data taken for about 22 years. Such an inquiry tries to characterize the local temperature variability and to link it with the global atmospheric oscillations.

## 2. Methodology and data

This section shortly describes the basis of the method adopted to elaborate the radiosounding data in the present analysis and the preliminary processing of the data aiming to create the input for the computation procedure.

### 2.1. Method

The principle component analysis is a powerful tool for examining the spatio-temporal variability of a scalar field presented by a physical variable [21, 22, 23]. The method requires the creation of the anomaly data matrix $\mathbf{F}$, or the matrix containing the deviations $\Delta V$ from the average trend of the variable $V$. The spatial distributions of the anomalies $\Delta V$ over a grid of $n$ observational points at certain time are presented as rows in $\mathbf{F}$, while the time series composed by $m$ measurements made at uniquely sampled times at corresponding grid points are given as columns:

$$\mathbf{F} = \begin{pmatrix} \Delta V_{\text{at point 1}}^{\text{at time } t_1} & \Delta V_{\text{at point 2}}^{\text{at time } t_1} & \cdots & \Delta V_{\text{at point } n}^{\text{at time } t_1} \\ \Delta V_{\text{at point 1}}^{\text{at time } t_2} & \Delta V_{\text{at point 2}}^{\text{at time } t_2} & \cdots & \Delta V_{\text{at point } n}^{\text{at time } t_2} \\ \vdots & \vdots & \cdots & \vdots \\ \Delta V_{\text{at point 1}}^{\text{at time } t_m} & \Delta V_{\text{at point 2}}^{\text{at time } t_m} & \cdots & \Delta V_{\text{at point } n}^{\text{at time } t_m} \end{pmatrix}. \quad (1)$$

The anomalies $\Delta V_{ij}$ $(i=1,2,\cdots,m; j=1,2,\cdots,n)$ are usually calculated by removing the corresponding trend $\overline{\mathbf{V}}_j$ from each column $\mathbf{V}_j$ of the matrix $\mathbf{F}'$, constructed similarly to $\mathbf{F}$ but containing the measurement data $V$ instead of $\Delta V$. The further step is to find the $n \times n$ covariance matrix $\mathbf{C} = [1/(m-1)]\mathbf{F}^T\mathbf{F}$, where $\mathbf{F}^T$ is the transposed matrix $\mathbf{F}$, which can be presented by solving the eigenvalue problem as:

$$\mathbf{CE} = \mathbf{E}\boldsymbol{\Lambda}. \quad (2)$$

The columns $(\mathbf{e_1} \quad \mathbf{e_2} \quad \cdots \quad \mathbf{e_n})$ of the matrix $\mathbf{E}$ are the eigenvectors of $\mathbf{C}$ and the diagonal matrix $\boldsymbol{\Lambda} = (\lambda_{jj} : \lambda_{11} \geq \lambda_{22} \geq \cdots \geq \lambda_{nn} \geq 0)$ presents the corresponding eigenvalues. Since the eigenvectors are orthogonal each other and result from field measurements data they are named Empirical Orthogonal Functions (EOF) and describe the spatial distribution of anomaly amplitude. Each of the values

$$w_j = \frac{\lambda_{jj}}{\text{Tr}(\boldsymbol{\Lambda})}, \quad (3)$$

where $\text{Tr}(\boldsymbol{\Lambda})$ is the sum of diagonal elements of matrix $\boldsymbol{\Lambda}$, represents the weight of the eigenvector $\mathbf{e_j}$, or the contribution of EOF$_j$ to the spatial distribution of $\Delta V$ represented by $\mathbf{C}$. In practice, only a few of EOFs have significant weight $w_j$ and hence, the decomposition performed through Eq. (2) projects the variations presented by $n$ vectors onto a space determined by $p$ orthogonal vectors ($p \ll n$). The cumulative weight $W_p = \sum_{j=1}^{p} w_j$ determined as the sum of the first $p$ values of $w_j$ in the most of the real cases rapidly increases to 1 (or 100%; see Fig. 3) and $p$ can be determined as the value of $j$ for which $W_p$ becomes reasonably close to 1.

The projection of the anomaly matrix $\mathbf{F}$ onto $j^{th}$ EOF

$$\mathbf{p_j} = \mathbf{F}\mathbf{e_j} \qquad (4)$$

is a vector named as the principal component (PC) of the corresponding EOF ($\mathbf{e_j}$) and characterizes the temporal EOF variations.

For the purposes of the present study $\mathbf{F}$ is structured so that each profile of temperature anomaly $\Delta T_{ij} = \Delta T_{t_i}(z_j)$ found for $z_j$ $(j=1, 2, \ldots, n)$ height levels provided by a radio sound launched at time $t_i$ $(i=1, 2, \ldots, m)$ is a row of the matrix:

$$\mathbf{F} = \begin{pmatrix} \Delta T_{11} & \Delta T_{12} & \cdots & \Delta T_{1n} \\ \Delta T_{21} & \Delta T_{22} & \cdots & \Delta T_{2n} \\ \vdots & \vdots & \cdots & \vdots \\ \Delta T_{m1} & \Delta T_{m2} & \cdots & \Delta T_{mn} \end{pmatrix}. \qquad (5)$$

Thus, each column is a time series characterizing the temperature anomalies at certain altitude level. In the present study, the EOFs and PCs have been found by means of the singular value decomposition of the anomaly matrix $\mathbf{F}$, an alternative method presenting $\mathbf{F}$ as a product of three matrices:

$$\mathbf{F} = \mathbf{P}\,\mathbf{S}\,\mathbf{E}^T, \qquad (6)$$

where PCs are the columns of the $m \times r$ matrix $\mathbf{P}$ and EOFs are the columns of the $r \times n$ matrix $\mathbf{E}$, respectively, and $r$ is the rank of $\mathbf{F}$, $r \leq \min(m,n)$. The diagonal matrix $\mathbf{S} = (s_{kk} : s_{11} \geq s_{22} \geq \cdots \geq s_{rr} \geq 0)$ contains the singular values, which are connected with the eiginvalues of the covariance matrix $\mathbf{C}$ as $s_{kk} = \sqrt{\lambda_{kk}}$. In addition, each EOF was multiplied by the associated singular value and each PC was divided by the same value, respectively that, according to Camp et al. [23] returns the EOFs in dimensional units, degree Kelvin in our case. The singular value decomposition was performed by using the corresponding MATLAB function and the spectral analysis of the PC components were made by applying the Lomb−Scargle [24, 25] periodogram approach.

*2.2 Data-set and preliminary elaboration*

The present study analyses the data provided by Vaisala radio sounds routinely launched at San Pietro Capofiume station (44°39'N, 11°36'E, 11 m amsl), located in the southeast part of the Po Valley, Italy, twice a day from August 1987 to March 2010. The radio sound gives the vertical distributions of the atmospheric pressure, temperature, relative humidity and, velocity and direction of the wind. Two types of radio sounds were used at the station: RS80, until mid September 2005 and RS92 after that, which are characterized by accuracy of the temperature measurements of ±0.2 and ±0.5 K, respectively [26]. The temperature sensor of RS80 is affected by lag error that was corrected through the procedure proposed by Tomasi et al. [26], while the sensor of RS92 does not need any corrections. Each radiosounding profile was projected by means of linear interpolation onto equally spaced height grid starting from the station surface level, 100 m as the second level and reaching 20 km after that with a step of 100 m. Figure 1 presents a result of this procedure applied on three temperature profiles observed at the station on different days of the examined period. Further, these profiles formed the matrix $\mathbf{F}'$, in which they were consecutively inserted as rows and the gaps in the columns of $\mathbf{F}'$, resulted from sporadically missing radio sound launches, were filled by means of linear interpolation to obtain sequences with a step of 12 hours. In order to eliminate the oscillations associated with micro- and meso-scales atmospheric processes considered shorter than 10 days [27], sequences, consisted of 10-day average values of the temperature as illustrated in Fig. 2 (a) by the red curve, replaced the columns in $\mathbf{F}'$. It is seen from Fig. 2 (a) that the temperature variations present well marked oscillations with a period close to 1 year modulated by lower frequency fluctuations. Since the amplitude of these two types of oscillations is quite different it seems reasonable to separate them for the next analysis. For that purpose, a running average procedure with 30-day window was applied to obtain the long-period variations given by the blue curve in Fig. 2 (a). The difference between sequences represented by the red and blue curves, shown in Fig. 2 (b), is considered to represent the intra-annual temperature oscillations, or the oscillations characterized by periods lower than 1 year that form the anomaly matrix $\mathbf{F}_1$ subject of the further analysis. Time series, presenting the long-period oscillations were detrended by removing from each of them the corresponding trend found through linear approximation that gives the azure curve in Fig. 2 (c). The spectral analysis showed that these oscillations were strongly dominated by the annual and semi-annual cycles that masked the other long-period fluctuations and hence, it was decided to remove the two cycles from the data. Such a filtering was performed by extracting the variations consisted of periods lower than 1 year, which are presented by the brown curve in Fig. 2 (c), from the long-period temperature oscillations given by the azure curve in the same figure. At the end, Fig. 2 (d) exhibits the resulting curve that represents the inter-annual oscillations composing the second anomaly matrix $\mathbf{F}_2$ analysed in the present study.

## 3. Results and discussion

Figure 3, which shows the cumulative weight $W_p$ of the EOFs obtained for the intra- and inter-annual temperature variations presented by matrices $\mathbf{F}_1$ and $\mathbf{F}_2$ respectively, indicates that in both cases the first four components explain 93 – 95% of the temperature variations. These EOFs are given in Fig. 4, while Figs. 5 and 6 exhibit the corresponding PCs on the left-hand side and their spectra on the right. The EOFs represent the vertical profiles of the temperature anomalies that can be considered the amplitude of variations determined by the corresponding PCs, so that each of the two groups containing four EOF – PC pairs defines the corresponding type of temperature variations, given by matrices $\mathbf{F}_1$ and $\mathbf{F}_2$, trough Eq. 6.

Figures 5 and 6 reveal that each PC in both cases is characterised by a spectral band dominating on the other frequencies. It can be seen from Fig. 5 (b) that the leading PC1 of the matrix $\mathbf{F}_1$ consists mainly of periods pertaining to 30 – 120 days band with peaks at 51, 54, 73 and 98 days, that can be associated with the MJO [4] and AO [6, 7] oscillations or with an interaction between them according to the conclusions made by Zhou and Miller [14]. In addition, these oscillations are modulated by the annual cycle that has secondary importance for the PC1 variations. The corresponding EOF1 given in Fig. 4 (a) shows that PC1 oscillations are presented in the troposphere between 1 and 8 km, while in the tropopause zone the amplitude decreases changing the sign that can be interpreted as changing of the fluctuation phase. In the low stratosphere the PC1 oscillations gradually decrease. It should be pointed out that this EOF1 – PC1 pair explains about 70% of the intra-annual temperature variations at the San Pietro Capofiume station.

The second component EOF2 of the anomaly matrix $\mathbf{F}_1$ that contributes the variations by about 12% is presented in Fig. 4 (b), while the corresponding PC2 and its spectrum are given in Figs. 5 (c) and (d), respectively. It is seen from Fig. 4 (b) that these oscillations are uniformly weak in all altitude range with a slight enhancement in the Upper Troposphere – Low Stratosphere (UTLS) range. The spectrum shows that the PC2 variations are strongly modulated by the annual and semi-annual cycles while the fluctuations with periods of 30-100 days are weaker by approximately an order of magnitude.

Figure 4 (c) shows EOF3 component that indicates almost null amplitude of the oscillations in the troposphere and values between -1 and 1 K in UPLS. This component has a low contribution (~7%) with a prevailing weight of the 120-day period in the corresponding PC3 component (Fig. 5 (e)) as the spectrum given in Fig. 5 (f) reveals. The EOF4 weight in the temperature oscillations is about 3% and this component presents variations mainly in the 9 – 15 km altitude range (see Fig. 4 (d)) that are consisted of periods between 30 and 120 days with peaks at 52, 62, 72, 82 and 120 days (Figs. 5 (g) and (h)).

Lower part of Fig. 4 and Fig. 6 exhibit the EOFs and PCs respectively, obtained from decomposition of the matrix $\mathbf{F}_2$, representing the temperature anomalies associated with the inter-annual oscillations. The first component EOF1 (Fig. 4 (e)) that contributes the oscillations by about 68%, shows a pattern analogous to the EOF1 of the intra-annual fluctuations (Fig. 4 (a)), presenting the largest amplitude of about 1 K in the troposphere between 1 and 9 km that rapidly decreases in the region of the tropopause. In the upper-altitude range the amplitude changes the sign and increases until about 13 km gradually decreasing to zero after that. The corresponding temporal variations presented by PC1, shown in Fig. 6 (a), are composed of spectral components ranging between 1- and 7-year periods with peaks at 1.3, 1.7, 2.3, 2.9 and 6.8 years, as Fig. 6 (b) indicates. The period at 2.9 years exhibits the highest spectral power followed by the periods at 1.7 and 6.8 years. The ratio between powers of 2.9- and 6.9-year periods is very similar to that presented by the corresponding components for the ESNO index spectrum given in the Stenseth et al. [10] study (their Fig. 2 (c)). Such a similarity leads to the assumption that the ENSO are strongly presented in the leading PC of the inter-annual temperature fluctuations. The other components that present a secondary importance could be associated with the QBO and its interaction with the annual cycle [8]. EOF2 exhibited in Fig. 4 (f) shows 18% contribution with negligible effect in the troposphere and increasing amplitude in UTLS. The temporal variations associated with this component, presented by PC2 in Fig. 6 (c), consist predominantly of 21-year cycle that can be linked up with the 22-year solar cycle and a secondary peak at 1.5 years as Fig. 6 (d) indicates. Similarly to EOF2, EOF3 presents higher contribution in UTLS (see Fig. 4 (g)) but only about 6% weight in the temperature variations. The oscillations of EOF3 (PC3) given in Fig. 6 (e) are dominated by 10.3-year cycle with secondary peaks at 1.9-, 2.3- and 2.9-year fluctuations. The first cycle could be accounted for the 11-year solar cycle, while the others can be interpreted as a manifestation of the QBO and ENSO or NAO fluctuation components. Figure 4 (h) shows that EOF4, which contributes the inter-annual temperature fluctuations by about 3%, shows significant amplitudes between 8 and 12 km, the region of the tropopause altitudes. The corresponding PC4 given in Fig. 6 (g) is composed by oscillations characterized by a large band between 4- and 8-year periods and a secondary peak at 1.5-year period. The oscillations pertaining to the first spectral range could be associated with the 6 – 10 year band of the NAO [12].

It is worth pointing out that all EOFs corresponding to the inter-annual temperature variations taken place at the station subject of the present study and given in the lower part of Fig. 4 exhibit more or less large amplitudes in the boundary layer.

## 4. Conclusions

The present study has analysed the temperature variations observed above the southeast part of the Po Valley, Italy up to 20 km by means of radio sounds launched twice a day during about 22-year period. The principle component analysis of the data allowed the projection of the temperature variations over four principle vertical distribution and temporal modes, respectively. Such an analysis shows that except of the annual cycle, strongly presented in all altitude levels, the intra-annual oscillations composed of periods between 30 and 120 days, together with the inter-annual variations with periods of 1 to 7 years predominantly determine the temperature fluctuations in the troposphere. The first type of oscillations can be associated with MJO and their interaction with AO, while the second group could be accounted for the ESNO and QBO. These variations vanish in the tropopause region and, changing the phase appear to different extent in the low stratosphere. The annual and semi-annual pulsations of the amplitude characterizing the short-period mode are of secondary importance and are presented mainly in the 10-13 km range, where the tropopase altitude usually varies, while the oscillations with 120-day period present minor contribution and could be observed in UTLS. The inter-annual oscillations described by EOF2-4 components that could be associated with global fluctuations as ESNO, QBO, NAO are negligibly weak in the troposphere and take place mainly in the UTLS region.

### Acknowledgements
The radiosoundings data recorded at San Pietro Capofiume station locted in the southeast part of the Po Valley have been downloaded from the website of the University of Wyoming (http://weather.uwyo.edu/upperair/sounding.html).

**References**

[1] G. C. Reid, "Seasonal and interannual temperature variations in the tropical stratosphere", *Journal of Geophysical Research*, vol. 99, no. D9, pp. 18923 – 18932, 1994

[2] X. Dou et al., "Seasonal oscillations of middle atmosphere temperature observed by Rayleigh lidars and their comparisons with TIMED/SABER observations", *Journal of Geophysical Research*, vol. 114, D20103, 2009.

[3] I. Isaia, "Oscillations and Cycles of Air Temperature in Russia", *Present environment and sustainable development*, vol. 8, nom. 1, 191 – 203, 2014.

[4] R. A. Madden and P. R. Julian, "Detection of a 40-50 day oscillation in the zonal wind in the tropical Pacific", *Journal of the Atmospheric Sciences*, vol. 28, pp. 702 – 708, 1971.

[5] Ch. Zhang, "Madden-Julian Oscillation" *Reviews of Geophysics*, vol. 43, RG2003, 2005

[6] W. J. Thompson and J. M. Wallace, "The Arctic Oscillation signature in the wintertime geopotential and temperature fields", *Geophysical Research Letters*, vol. 25, no. 9, pp. 1297 – 1300, 1998.

[7] M. P. Baldwin and T. J. Dunkerton, "Propagation of the Arctic Oscillation from the stratosphere to the troposphere", *Journal of Geophysical Research*, vol. 104, no. D24, pp. 30937 – 30946, 1999.

[8] M. P. Baldwin et al., "The Quasi-biennial oscillation", Reviews of Geophysics, vol. 39, no. 2, pp. 179–229, 2001.

[9] C. Wang, C. Deser, J.-Y. Yu, P. DiNezio, A. Clement, "El Niño and Southern Oscillation (ENSO): A Review", in *Coral Reefs of the Eastern Pacific*, pp. 3 – 19, 2012.

[10] N. Chr. Stenseth, G. Ottersen, J. W. Hurrell, A. Mysterud, M. Lima, K.-S. Chan, N. G. Yoccoz and B. Ådlandsvik, "Studying climate effects on ecology through the use of climate indices: the North Atlantic Oscillation, El Niño Southern Oscillation and beyond (Review)", *Proceedings of the Royal Society of London B*, vol. 270, pp. 2087–2096, 2003

[11] R. García-Herrera, N. Calvo, R. R. Garcia and M. A. Giorgetta, "Propagation of ENSO temperature signals into the middle atmosphere: A comparison of two general circulation models and ERA-40 reanalysis data", Journal of Geophysical Research, vol. 111, D06101, 2006.

[12] H. Wanner et al., "North Atlantic Oscillation – concepts and studies", *Surveys in Geophysics*, vol. 22, pp. 321–382, 2001.

[13] N.-C. Lau, "Interactions between Global SST Anomalies and the Midlatitude Atmospheric Circulation", Bulletin of the American Meteorological Society, vol. 78, nom. 1, 21 - 33, 1997.

[14] S. Zhou and A. J. Miller, "The interaction of the Madden–Julian Oscillation and the Arctic Oscillation", *Journal of climate*, vol. 18, pp. 143 – 159, 2005.

[15] M. F. Stuecker, A. Timmermann, F.-F. Jin, S. McGregor and H.-L. Ren, "A combination mode of the annual cycle and the El Niño/Southern Oscillation". Nature Geosciences, vol. 6, pp. 540 – 544, 2013.

[16] B. Dewitte, C. Cibot, C. Périgaud, S.-I. An, L. Terray, "Interaction between Near-Annual and ENSO modes in a CGCM simulation: role of the Equatorial background mean state", Journal of climate, vol. 20, pp. 1035 – 1052, 2007.

[17] X. Zheng and R. E. Basher, "Structural time series models and trend detection in global and regional temperature series", *Journal of climate*, vol. 12, pp. 2347 – 2358, 1999.

[18] A. J. Simmons, K. M. Willett, P. D. Jones, P. W. Thorne, and D. P. Dee, "Low-frequency variations in surface atmospheric humidity, temperature, and precipitation: Inferences from reanalyses and monthly gridded observational data sets", *Journal of Geophysical Research*, 115, D01110, 2010

[19] D. T. Mihailović, G. Mimić, and I. Arsenić, "Climate Predictions: The Chaos and Complexity in Climate Models", *Advances in Meteorology*, vol. 2014, Article ID 878249, 14 p., 2014.

[20] S. A. Solman, "Regional Climate Modeling over South America: A Review", *Advances in Meteorology*, vol. 2013, Article ID 504357, 13 p., 2013.

[21] H. Björnsson and S. A Venegas, "A manual for EOF and SVD analyses of climatic data", Report № 97-1, Montreal, Quebec, 1997.

[22] A. Hannachi, I. T. Jolliffe and D. B. Stephenson, "Empirical orthogonal functions and related techniques in atmospheric science: A review", *International Journal of Climatology*, vol. 27, pp. 1119–1152, 2007.

[23] C. D. Camp, M. S. Roulston, and Y. L. Yung, "Temporal and spatial patterns of the interannual variability of total ozone in the tropics", *Journal of Geophysical Research*, vol. 108(D20), 4643, 2003.

[24] N. R. Lomb, "Least-squares frequency analysis of unequally spaced data", *Astrophysics and Space Sciences*, vol. 39, pp.447–462, 1976.

[25] J. D. Scargle, "Studies in astronomical time series analysis. II. Statistical aspects of spectral analysis of unevenly spaced data", *Astrophysical Journal*, vol. 263, pp.835–853, 1982.

[26] C. Tomasi, B. Petkov, E. Benedetti, V. Vitale, A. Pellegrini, G. Dargaud, L. De Silvestri, P. Grigioni, E. Fossat, W. L. Roth and L. Valenziano, "Characterization of the atmospheric temperature and moisture conditions above Dome C (Antarctica) during austral summer and fall months", *Journal of Geophysical Research*, vol. 111, D20305, 2006

[27] D. G. Steyn, T.R. Oke, J. E. Hay and J. L. Knox, "On scales in meteorology and climatology." Climate Bulletin, vol. 39, pp. Climate bulletin 1-8, 1981.

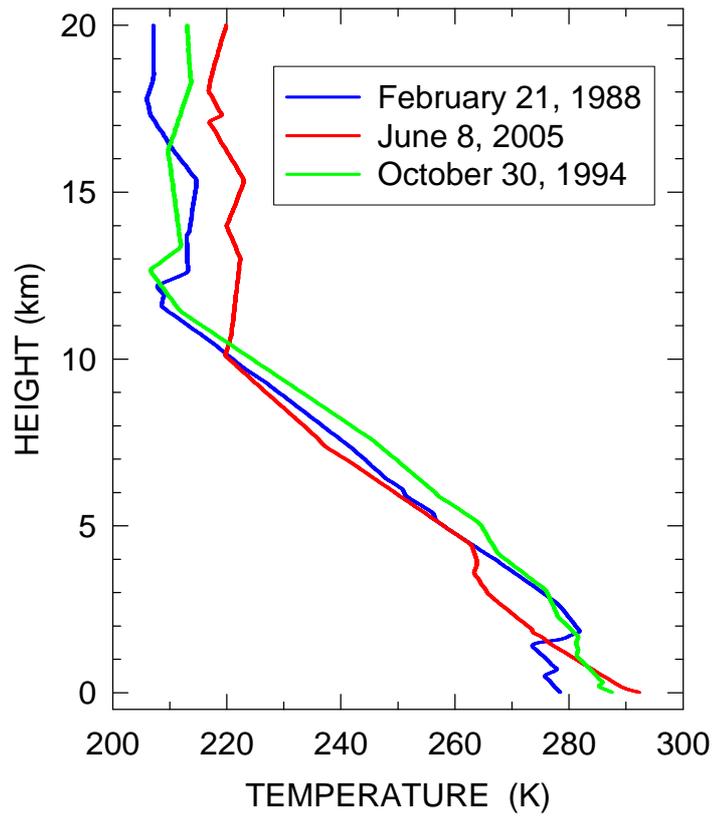

**Fig. 1** Three temperature profiles recorded at the San Pietro Capofiume station on different days indicated in the graph.

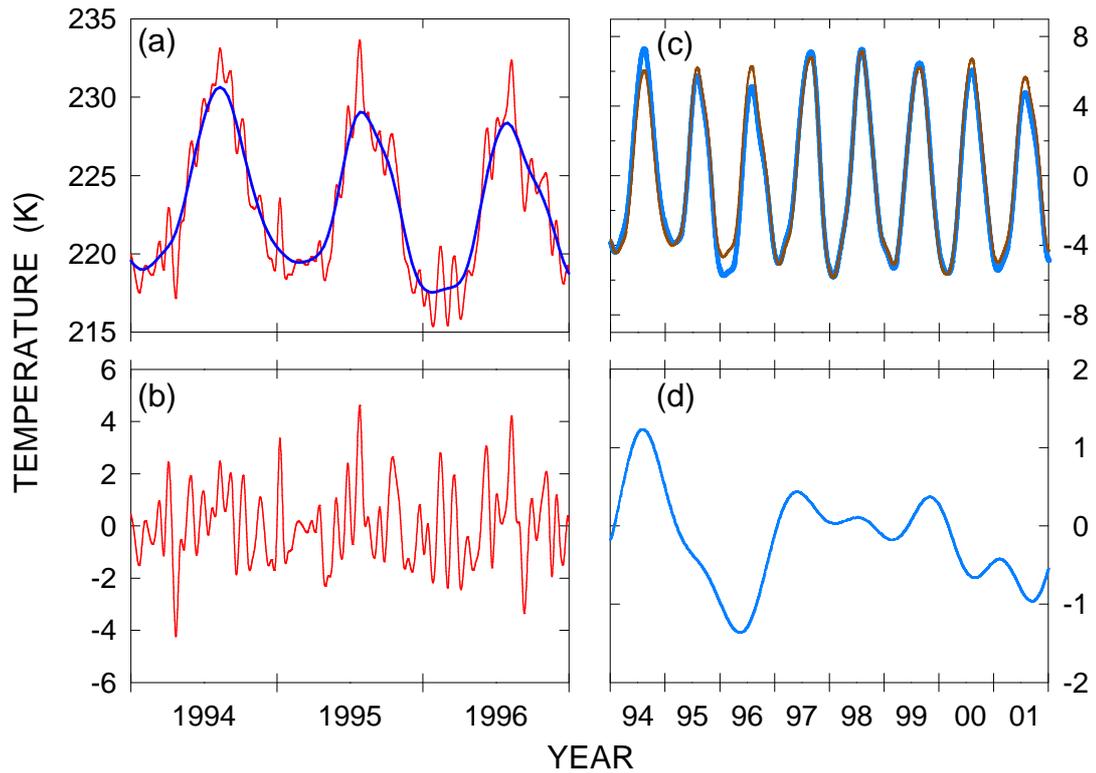

Fig. 2 Illustration of the preliminary data processing as it was applied on the data at 10 km altitude. Panel (a) presents the 10-day averaged temperature data (red curve) and the corresponding long-period march (blue curve), while the difference between two curves that is considered to represent the intra-annual variations, is given in panel (b). Panel (c) shows detrended long-period variations presented by the azure curve and the oscillations, composed by lower than 1 year cycles (brown curve). The difference between them, given in panel (c) can be associated with the inter-annual variations.

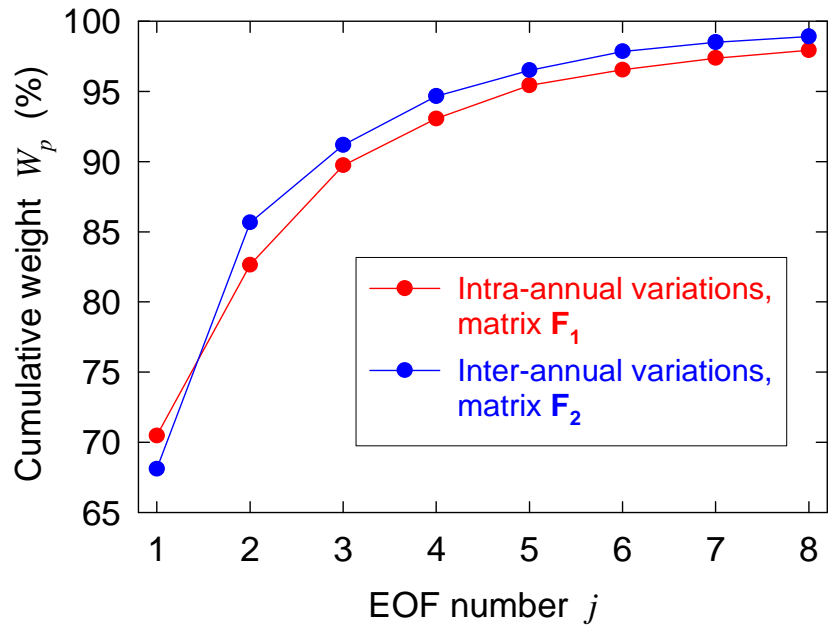

Fig. 3 Cumulative weight $W_p$ (in percents) of the EOFs found for both cases of temperature anomalies, subject of the present study.

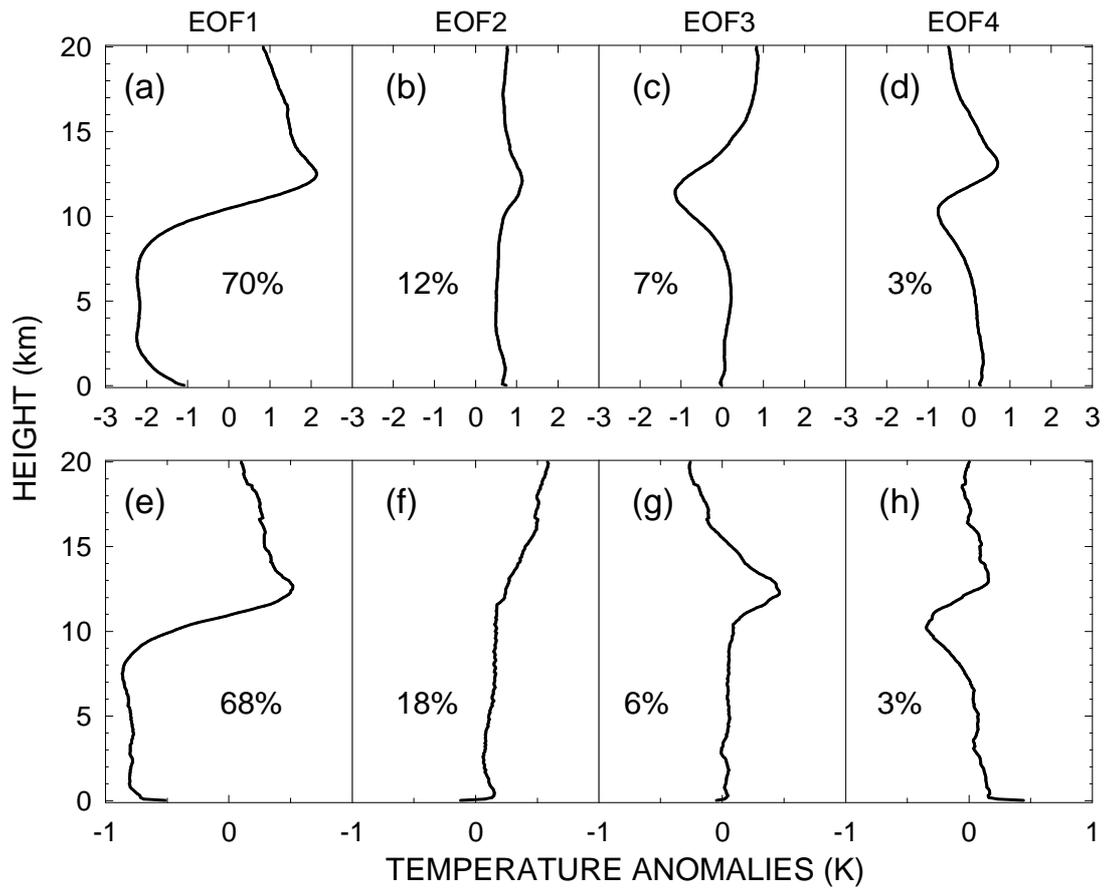

Fig.4. The first four EOFs evaluated for the matrix $\mathbf{F}_1$ (up) and $\mathbf{F}_2$ (down). The corresponding weight $w_j$ of each component, evaluated through Eq. (3) is given in percents in each panel.

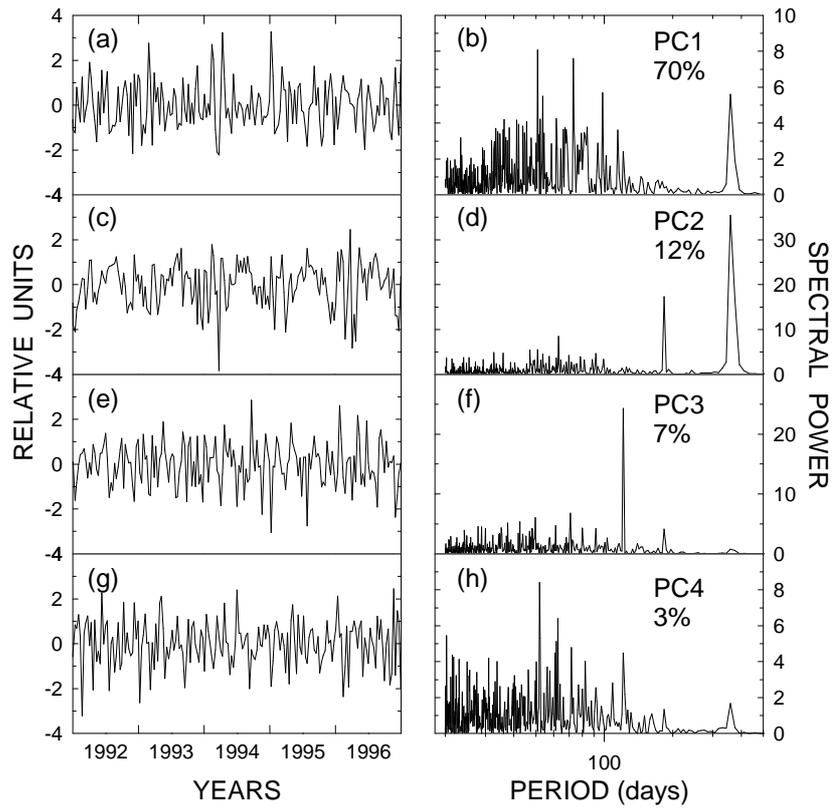

Fig. 5. The leading four PCs (left, presented for a 5-year period) associated with the corresponding EOFs found from the decomposition of matrix $\mathbf{F}_1$ (see upper part of Fig. 4) together with their spectra (right). The two panels of each row correspond to the same PC indicated on the right, together with the weight $w_j$ in percent.

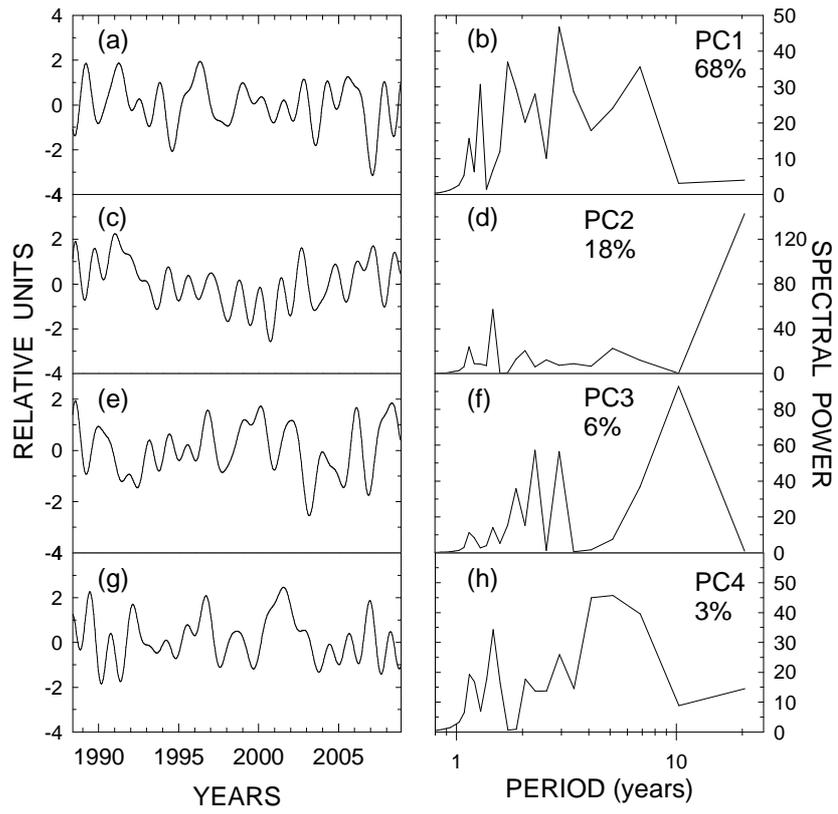

Fig. 6. The same as Fig. 5 but for the PCs of matrix $\mathbf{F}_2$ corresponding to EOFs presented in the lower part of Fig. 4.